# First-principles study on Pr-doped Bilayer Nickelate $La_3Ni_2O_7$

Zihao Huo[1], Peng Zhang[2], Haoliang Shi[1], Xiaochun Yan[1], Defang Duan[1,*], Tian Cui[3,1]

[1]*Key Laboratory of Material Simulation Methods & Software of Ministry of Education, State Key Laboratory of Superhard Materials, College of Physics, Jilin University, Changchun 130012, China*

[2]*MOE Key Laboratory for Non-equilibrium Synthesis and Modulation of Condensed Matter, Shaanxi Province Key Laboratory of Advanced Functional Materials and Mesoscopic Physics, School of Physics, Xi'an Jiaotong University, Xi'an 710049, China*

[3]*Institute of High Pressure Physics, School of Physical Science and Technology, Ningbo University, Ningbo 315211, China*

*Corresponding author: duandf@jlu.edu.cn

**Abstract:** Recently, the Pr-doped Ruddlesden-Popper phase of bilayer nickelate $La_3Ni_2O_7$ has been reported to exhibit a superconducting transition temperature (Tc) of 82.5 K and superconducting volume fraction of about 57 % at high pressure. However, the effect of Pr-doping on $La_3Ni_2O_7$ remains unclear. Here, we studied the crystal structures and electronic properties of Pr-doped $La_3Ni_2O_7$ at 0 and 15 GPa based on the first-principles calculations to explore the doping effect of Pr. Our findings indicate that the praseodymium atoms prefer to occupy the outer La-O layers site. We then investigated the evolution of crystal structures in both the ambient pressure phase and high pressure phase of $La_3Ni_2O_7$ as a function of doping concentration, revealing inconsistent trends in their evolution with increasing doping levels. Finally, by fitting the bilayer two-orbital model, we propose that doping Pr may benefit for superconductivity of $La_3Ni_2O_7$. These results not only can help the further experimental search of RP phase nickelate at lower pressure, but also provide helpful guide for understanding the effect of chemical pressure in isovalent doped RP phase nickelate superconductor.

Superconductivity at $T_c$ = 80 K has recently been reported above 14 GPa in the bilayer Ruddlesden-Popper (RP) phase of La$_3$Ni$_2$O$_7$ [1-5], which thus introduces a new family of high-temperature superconductors after discovery of cuprates [6-9], iron-based [10-15], and infinite-layer nickelates [16-20]. And a 'right-triangle' shape of the superconducting region has been observed in the pressure-temperature phase diagram [21]. In which the emergence of superconductivity is accompanied by a structural transition from the *Amam* to *I*4/*mmm* phase under pressure.

This breakthrough has attracted extensive investigation on the electronic properties and superconductivity mechanism in pressured La$_3$Ni$_2$O$_7$ [22-36]. Density functional theory (DFT) calculations show that the Ni-$d_{z^2}$ orbitals can be seen as near half-filled and more localized, whereas the Ni-$d_{x^2-y^2}$ orbitals is near quarter-filling and more itinerant [25, 37-40]. The electronic correlation of Ni-$d$ orbitals has been experimental measured at ambient pressure [41-43] and are in lined with the previous DFT calculations. Besides, a minimal bilayer two-orbital model has been proposed to captures the key ingredients of La$_3$Ni$_2$O$_7$ [38, 44]. Both the Ni-$d_{z^2}$ orbitals and the Ni-$d_{x^2-y^2}$ orbitals have been suggested to play an important role in low-energy physics. In the following study, the vertical hopping term $t_\perp$ of Ni-$d_{z^2}$ orbitals and intralayer hybridization $V$ between Ni-$d_{z^2}$ and Ni-$d_{x^2-y^2}$ orbitals have been suggested as the key factors for the superconductivity in La$_3$Ni$_2$O$_7$ [44-49]. For example, $t_\perp$ and $V$ were linked to the pairing phase coherence and the pairing glue based on cluster dynamical mean-field theory study [46].

The effect of replacing La with other rare-earth elements have been widely explored. The $T_c$ of A$_3$Ni$_2$O$_7$ can be expected to decrease with decreasing radius of the rare-earth ion by using random phase approximation method [39]. DFT calculations show that Tb$_3$Ni$_2$O$_7$ has been suggested to show superconductivity at ambient pressures due to its higher density of states (DOS) at the Fermi level ($E_F$) [50]. Recently, high pressure experimental suggested that replacing La$^{3+}$ with smaller rare-earth A$^{3+}$ (A = Pr, Nd, Tb) could introducing chemical pressure into La$_3$Ni$_2$O$_7$ and affect its critical pressure for structural transition [51]. Notably, Pr-doped *I*4/*mmm* phase of $La_2PrNi_2O_7$ has been observed with the onset $T_c$ of 82.5 K at pressure of about 18 GPa, which is slightly higher than the *I*4/*mmm* phase of $La_3Ni_2O_7$ [52]. Besides, the zero-resistance $T_c$ of 60 K and superconducting volume fraction of ~ 57% provided convincing evidence for the bulk high-temperature superconductivity in RP phase nickelate. Therefore, doping praseodymium can significantly improve the superconductivity of $La_3Ni_2O_7$. However, a microscopic understanding of Pr-doping effect is yet to be understood.

Here, we performed a comprehensive first-principles study on the crystal structures and electronic properties of Pr-doped $La_3Ni_2O_7$ at 0 and 15 GPa. Firstly, we calculated the enthalpies of the inner and outer La-O layers doping site and suggested that the Pr atoms would prefer to occupy the outer layers site. We then investigated the evolution of the lattice parameters as a function of the Pr doping concentration and verified the validity of the structural model we proposed. After that, we discussed the influence of the doping effects on the superconductivity of $La_3Ni_2O_7$ by fitting the bilayer two-orbital model and proposed that doping Pr into $La_3Ni_2O_7$ may benefited to the superconductivity.

The $La_3Ni_2O_7$ has an orthorhombic *Amam* symmetry at ambient pressure, which can be termed as an inter-growth of two planes of $NiO_6$ octahedra planes and La-O fluorite-type layers stacking along the *c* directions. We define the upper and lower La-O layers as the outer layers and the middle La-O layers as the inner layers (Fig. 1(a)), which is similar as the definition of Ni-O layers in trilayer RP phase $La_4Ni_3O_{10}$ [53]. We then investigated the doping effect on Pr-doped $La_3Ni_2O_7$ using the supercell method. Fig. 1(c) show the results for the enthalpies difference of $La_{12-m}Pr_mNi_8O_{28}$ compared to all the Pr atoms occupy at outer layers at ambient pressure. We can find that $La_{12-m}Pr_mNi_8O_{28}$ (m = 1-4) has the lowest enthalpies when all the praseodymium atoms occupying the outer layers site. When the number of praseodymium atoms at inner layers site increases, the enthalpy of $La_{12-m}Pr_mNi_8O_{28}$ increases. Therefore, all the praseodymium atoms occupying the outer layers site in $La_{12-m}Pr_mNi_8O_{28}$ is enthalpy preferred. At the pressure of 15 GPa, $La_3Ni_2O_7$ adopts a tetragonal symmetry (space group of *I*4/*mmm*). Similar with the *Amam* phase, we can also define the La-O layers in *I*4/*mmm* phase of La$_3$Ni$_2$O$_7$ as outer and inner layers (Fig. 1(b)). We also calculated the enthalpies of La$_{12-m}$Pr$_m$Ni$_8$O$_{28}$ at 15 GPa and plotted in Fig. 1(d). The results show that the relationships between the enthalpies of La$_{12-m}$Pr$_m$Ni$_8$O$_{28}$ and the number of inner layers site occupancies are similar with those of at ambient pressure. Therefore, the praseodymium atoms occupying the outer layers site in La$_{12-m}$Pr$_m$Ni$_8$O$_{28}$ is enthalpy preferred. This phenomenon may due to the higher vacancy formation energy of outer apical oxygen [54], which can withstand the effect of chemical pressure when doping Pr atoms. We only focused on the structures of La$_{12-m}$Pr$_m$Ni$_8$O$_{28}$ that all the praseodymium atoms occupied the outer layers site in the following part.

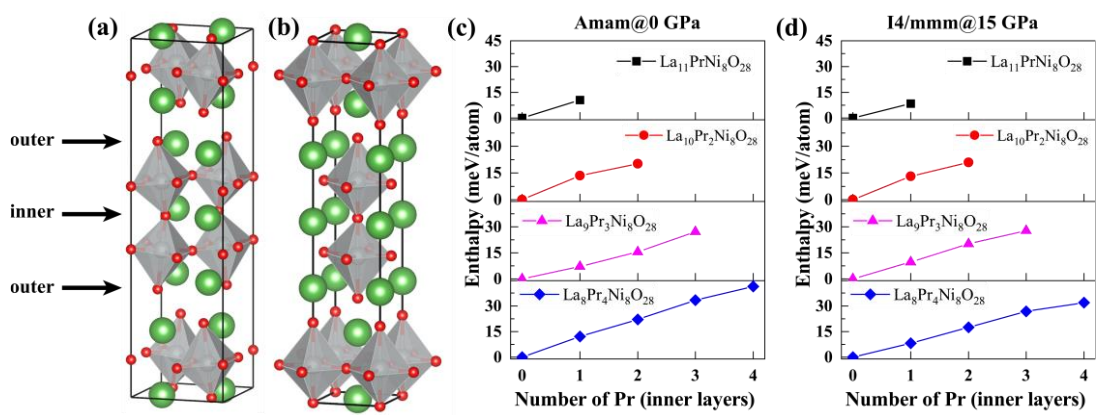

FIG 1. (a-b) Crystal structures of the *Amam* and *I*4/*mmm* phase of La$_3$Ni$_2$O$_7$. The green, grey, and red spheres represent La, Ni, and O atoms, respectively. (c-d) Calculated enthalpies of *Amam* and *I*4/*mmm* phase as a function of the number of inner layers site occupancies at 0 and 15 GPa, respectively.

Fig. 2 represented the crystal structures of Pr-doped *Amam* and *I*4/*mmm* phase of La$_3$Ni$_2$O$_7$. For Pr-doped *Amam* phase of La$_3$Ni$_2$O$_7$ at ambient pressure, the Pr would random replaced one of the La in outer layers in *Pm* phase of La$_{11}$PrNi$_8$O$_{28}$ due to the same Wyckoff position of La in outer layers (8g). In *Pma*2 phase of La$_{10}$Pr$_2$Ni$_8$O$_{28}$, the Pr atoms would replace the same La-O layer, with ABAABC stacking along c axis. For the *Pm* phase of La$_9$Pr$_3$Ni$_8$O$_{28}$, the stacking sequence is ABACBD along c axis. While the layer stacking for *Ama*2 phase of La$_8$Pr$_4$Ni$_8$O$_{28}$ is ABCABC along c axis. For Pr-doped *I*4/*mmm* phase of La$_3$Ni$_2$O$_7$ at 15 GPa, the doping site of Pr is similar with that of at ambient pressure in La$_{11}$PrNi$_8$O$_{28}$ and La$_{10}$Pr$_2$Ni$_8$O$_{28}$. While the *P*4*mm* phase of La$_9$Pr$_3$Ni$_8$O$_{28}$ forming a ABCABD stacking sequence along c axis. And in the *P4mm* phase of $La_8Pr_4Ni_8O_{28}$, the layer stacking along c axis is ABCDBD. Therefore, structural transition would rearrange the doping site of Pr atoms when the doping concentration is higher. Notably, the structure of $NiO_6$ octahedra maintained in all of these crystal structures and suggested that it is very robust when doping Pr atoms. It is reported that the $NiO_6$ octahedra plays an important role in the superconductivity of $La_3Ni_2O_7$ [40] and its robustness may explain the maintain of superconductivity in Pr-doped $La_3Ni_2O_7$ [52].

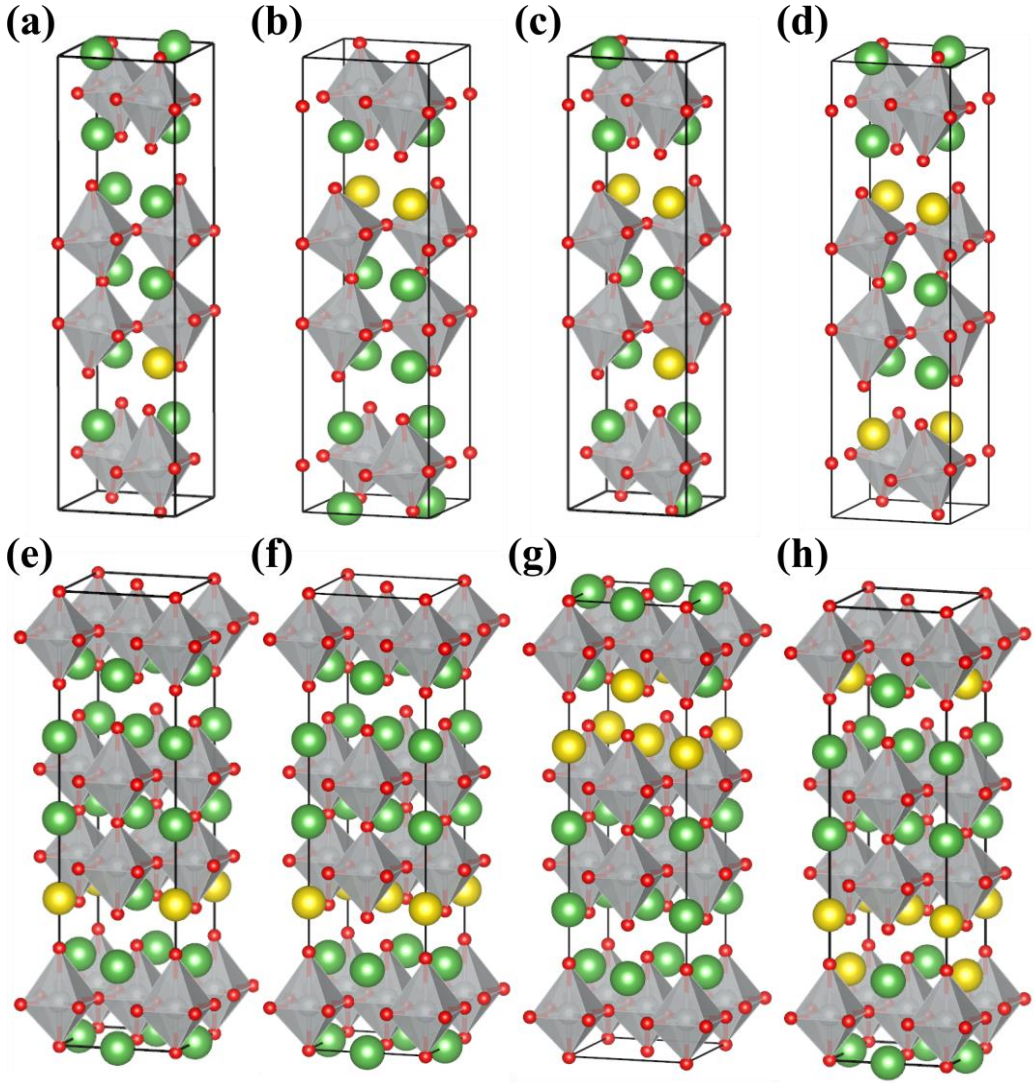

FIG. 2. Crystal structures of (a) *Pm*-$La_{11}PrNi_8O_{28}$, (b) *Pma*2-$La_{10}Pr_2Ni_8O_{28}$, (c) *Pm*-$La_9Pr_3Ni_8O_{28}$, and (d) *Ama*2-$La_8Pr_4Ni_8O_{28}$. These crystal structures are showed in the framework of *Amam*-$La_3Ni_2O_7$. Crystal structures of (e) *P4mm*-$La_{11}PrNi_8O_{28}$, (f) *P4mm*-$La_{10}Pr_2Ni_8O_{28}$, (g) *P4mm*-$La_9Pr_3Ni_8O_{28}$, and (h) *P4mm*-$La_8Pr_4Ni_8O_{28}$. These crystal structures are showed in the framework of *I*4/*mmm*-$La_3Ni_2O_7$. The green, yellow, grey, and red spheres represent La, Pr, Ni, and O atoms, respectively.

We then represented the lattice parameters and average apical Ni-O-Ni bond angle as a function of Pr-doping concentration in Fig. 3. At ambient pressure, as the Pr atoms increases from 0 to 4, the value of a decreases from 5.369 to 5.319 Å and the value of c decreases from 20.642 to 20.454 Å, while the value of b increases from 5.442 to 5.488 Å. And the volume decreases from 603 to 597 Å$^3$. Notably, the value of average apical Ni-O-Ni bond angle possess a decreased trend and decreases from 167.4 to 160.8°. And this phenomenon has been observed in high-pressure experiment [51, 52]. Besides, the lattice parameters decreased and the apical Ni-O-Ni bond angle increased when introducing external pressure into *Amam*-La$_3$Ni$_2$O$_7$ [50]. Which means the influence of chemical pressure and external pressure on the crystal structure of *Amam*-La$_3$Ni$_2$O$_7$ is different. When the pressure is 15 GPa, due to the structural symmetry confined, the value of a equals to b in the whole doping concentration we studied. When the m increases, both the a (b) and c decreased from 5.249 and 19.988 to 5.237 and 19.713 Å, respectively, and the volume decreased from 550.7 to 540.6 Å$^3$. While the apical Ni-O-Ni bond angle unchanged and maintained 180°. It is interested that the evolution of lattice parameters and apical Ni-O-Ni bond angle when increasing the doping concentration is similar with that of increasing pressure into *I*4/*mmm*-La$_3$Ni$_2$O$_7$ [50]. It means that the chemical pressure plays a similar role on the crystal structures of *I*4/*mmm*-La$_3$Ni$_2$O$_7$ compared with the external pressure.

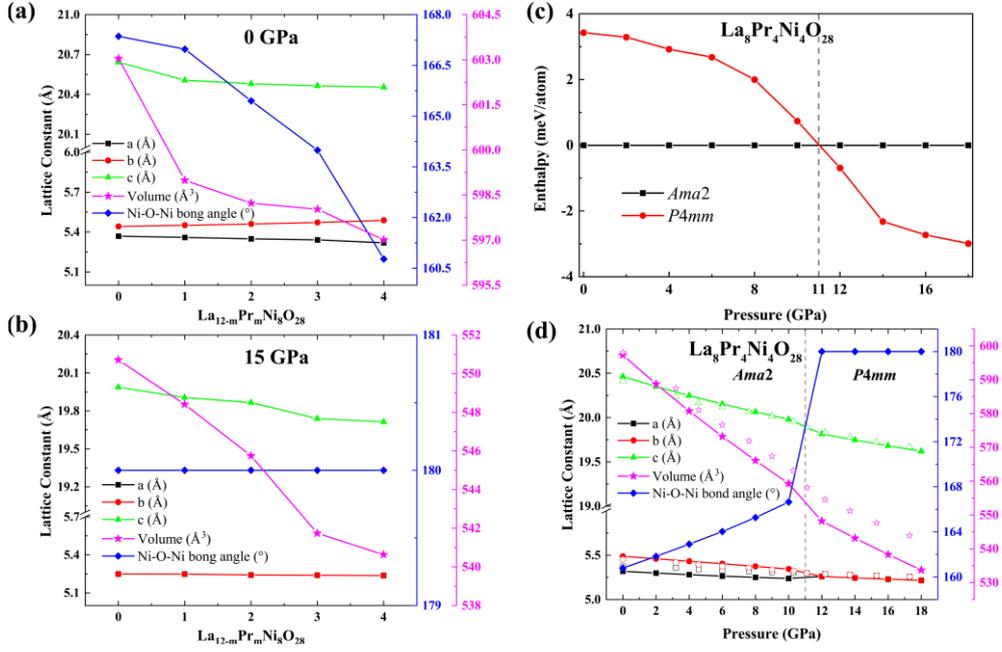

FIG 3. Calculated lattice parameters, volume, and average apical Ni-O-Ni bond angle as a function of Pr-doping concentration at (a) ambient pressure and (b) 15 GPa. (c) Calculated enthalpies curves for *Ama*2 and *P4mm* phase of $La_8Pr_4Ni_4O_{28}$ as a function of pressure. (d) Evolution of lattice parameters, volume, and average apical Ni-O-Ni bond angle of $La_8Pr_4Ni_4O_{28}$ as a function of pressure. The solid and open symbols correspond to the data of our calculation results and Wang *et al*.' experiment [52].

Enthalpy difference curves for *Ama*2 and *P4mm* phase of $La_8Pr_4Ni_4O_{28}$ have been presented in Fig. 3(c). *Ama*2 phase is the most stable structure when the pressure is lower than 11 GPa. When the pressure is above 11 GPa, the stable phase of $La_8Pr_4Ni_4O_{28}$ undergoes a transition from *Ama*2 to *P4mm* symmetry. It means that the critical pressure for structural transition from *Ama*2 to *P4mm* phase is 11 GPa, which is closely agree with that of 11.1 GPa for the high pressure experiment [52]. We also presented the evolution of lattice parameters, volume, and average apical Ni-O-Ni bond angle of $La_8Pr_4Ni_4O_{28}$ as a function of pressure by solid symbols in Fig. 3(d). For comparison, we overlay on the experimentally measured lattice parameters by the open symbols. For *Ama*2 phase, it can be observed that the experimentally measured lattice parameters *a* is slightly larger than our calculated value, whereas the experimentally measured lattice parameter *b* is in good agreement with our calculation at pressures. For *P4mm* phase, the experimentally measured lattice parameters *a* (*b*) is consistently slightly larger than our calculated results. For the lattice parameter $c$, the experimentally measurements are in excellent agreement with our calculated results in the whole pressure range we studied. Besides, the volume of the experiment measured are slight larger than our results. These results suggested that the structure model we employed can accurately describe the crystal structure of Pr-doped $La_3Ni_2O_7$. It is worth to note that the vertical Ni-O-Ni bond angle in the $Ama2$ phase of $La_8Pr_4Ni_4O_{28}$ gradually increased from 160.8 to 166.6° with the pressure increased from 0 to 10 GPa. However, when the structural transition occurred, the vertical Ni-O-Ni bond angle changed abruptly from 166.6 to 180°. This suggested that the pressure-induced structural transition is the main reason to change the vertical Ni-O-Ni bond angle to 180° for Pr-doped $La_3Ni_2O_7$.

Next, we calculated the electronic structures of Pr-doped $La_3Ni_2O_7$, and shown the results of $La_3Ni_2O_7$, $La_{11}PrNi_8O_{28}$, and $La_8Pr_4Ni_8O_{28}$ in Fig. 4. The electronic structures of $La_{10}Pr_2Ni_8O_{28}$ and $La_9Pr_3Ni_8O_{28}$ have been plotted in Fig. S1-4 of Supplemental Material (SM). For $Amam$ phase of $La_3Ni_2O_7$ at ambient pressure, the Ni-$d_{x^2-y^2}$ and Ni-$d_{z^2}$ orbitals are well separated from the Ni-$t_{2g}$ orbitals, and only the Ni-$d_{x^2-y^2}$ and O-$p$ orbitals dominate across the $E_F$ (Fig. 4(a)), in line with previous works [1, 40, 42, 50]. When doping Pr into $La_3Ni_2O_7$, the Pr-$f$ orbitals can be observed around ~ -2 and 3 eV and do not contribute to the $E_F$ due to the strong electronic correlated interaction (see Fig. S5 of SM). The splitting energy of two Ni-$d_{x^2-y^2}$ orbitals around -0.8 eV can describe the interlayer coupling strength of Ni-$d_{x^2-y^2}$ orbitals [38], and it increases from 0.01 eV in $Amam$-$La_3Ni_2O_7$ to 0.12 eV in $Ama2$-$La_8Pr_4Ni_8O_{28}$. Therefore, Pr-doped $Amam$-$La_3Ni_2O_7$ would significantly increasing the interlayer coupling strength of Ni-$d_{x^2-y^2}$ orbitals and enhanced the interlayer coupling of the $NiO_2$ plane. Besides, the energy of the flat bonding state of Ni-$d_{z^2}$ orbitals around Γ point are increased from -83.7 meV ($Amam$-$La_3Ni_2O_7$) to -43 meV ($Ama2$-$La_8Pr_4Ni_8O_{28}$) when the doping concentration increasing. It means that doping Pr atoms would upward shifts the two proximal flat bands of Ni-$d_{z^2}$ orbitals around Γ point to the $E_F$. It is noticed that the metallization of the Ni-$d_{z^2}$ orbitals play an important role in the superconductivity of $La_3Ni_2O_7$ [1, 40]. And the increment of the flat bonding state energy of Ni-$d_{z^2}$ orbitals may reduce the critical pressure of the superconductivity in $La_3Ni_2O_7$.

At 15 GPa, the Ni-$d_{z^2}$ bonding states lift upwards crossing the $E_F$ and a small hole Fermi pocket emerges around the Γ point. When doping praseodymium, the energy of Ni-$d_{z^2}$ orbitals around Γ point increased from 8 to 32.8 meV, while the energy of Ni-$d_{x^2-y^2}$ orbitals around Γ point decreased from -1.09 to -1.11 eV. It means that some electrons have been transformed from Ni-$d_{x^2-y^2}$ to Ni-$d_{z^2}$ orbitals and the self-doping effect occurs between these two orbitals. Besides, the splitting energy of two Ni-$d_{x^2-y^2}$ orbitals around -1.1 eV increased from 36 meV (*I*4/*mmm*-$La_3Ni_2O_7$) to 47 meV (*P*4*mm*-$La_8Pr_4Ni_8O_{28}$). Therefore, doping Pr into *I*4/*mmm*-$La_3Ni_2O_7$ would slightly enhance the interlayer coupling strength of Ni-$d_{x^2-y^2}$ orbitals.

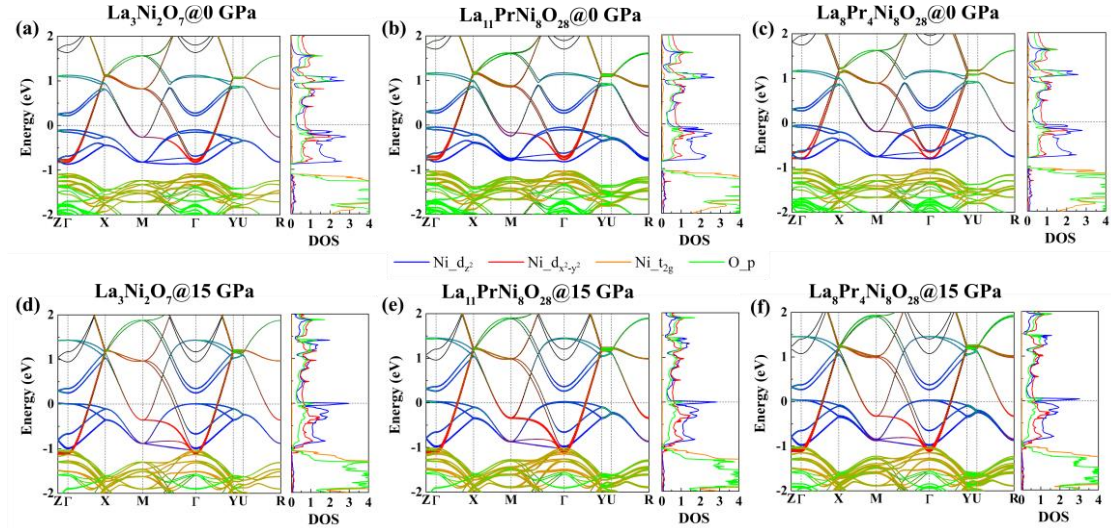

FIG 4. Projected electronic band structures of Ni cations and O anions of (a) *Amam* phase of $La_3Ni_2O_7$ at ambient pressure, (b) *Pm* phase of $La_{11}PrNi_8O_{28}$ at ambient pressure, (c) *Ama*2 phase of $La_8Pr_4Ni_8O_{28}$ at ambient pressure, (d) *I*4/*mmm* phase of $La_3Ni_2O_7$ at 15 GPa, (e) *P*4*mm* phase of $La_{11}PrNi_8O_{28}$ at 15 GPa, and (f) *P*4*mm* phase of $La_8Pr_4Ni_8O_{28}$ at 15 GPa. The corresponding DOS (states/eV/Ni) are shown on the right. The horizontal grey dash line means the Fermi level.

We then calculated the evolution of Ni-$d_{x^2-y^2}$, Ni-$d_{z^2}$ DOS of $La_8Pr_4Ni_4O_{28}$ at $E_F$, and $T_c$ (obtained by the previous research [52]) as a function of pressure, as shown in Fig. S6 of SM. The $T_c$ of $La_8Pr_4Ni_4O_{28}$ emergencies at about 8 GPa, and reveal with a maximum $T_c$ of 82.5 K at 16 GPa. Then the superconductivity is suppressed when applying a higher pressure. Interestingly, when the pressure below 10 GPa, the Ni-$d_{z^2}$ DOS increases slightly with the increasing pressure. It increases rapidly with the increasement of pressure at pressure above 10 GPa, and reaches its maximum at 16 GPa. Then it decreases with increasing pressure. While the Ni-$d_{x^2-y^2}$ DOS slightly decreases at the pressure range we studied. It means that the $T_c$ of $La_8Pr_4Ni_4O_{28}$ is close related to the Ni-$d_{z^2}$ DOS at $E_F$. Therefore, the Ni-$d_{z^2}$ orbitals significantly affect the superconductivity of $La_8Pr_4Ni_4O_{28}$, and it implies that $La_8Pr_4Ni_4O_{28}$ may obtained a $s_\pm$ pair symmetry similar to $La_3Ni_2O_7$ [24, 30, 33, 55].

Since the symmetry breaking occurred when doping Pr into $La_3Ni_2O_7$, $La_{12-m}Pr_mNi_8O_{28}$ exists the band folding phenomenon. To gain insight into the electronic properties of the Pr-doped $La_3Ni_2O_7$, we used the effective band structure method [56, 57] to calculated the unfolded electronic structures of $La_{12-m}Pr_mNi_8O_{28}$ and fitted out DFT unfolded band structure to the bilayer two-orbitals model as proposed by Luo *et al*. [38]. The unfolded band structures and the band structures of the bilayer two-orbital model have been represented in Fig. S7-14 of SM and the tight-binding parameters have been listed in Table S1-2 of SM. Fig. 5 plotted part of the tight-binding parameters as a function of Pr-doping concentration. At ambient pressure, among the out-of-plane hopping between two Ni-$d_{x^2-y^2}$ orbitals $t_\perp^x$, out-of-plane hopping between two Ni-$d_{z^2}$ orbitals $t_\perp^z$, and intralayer hopping between Ni-$d_{x^2-y^2}$ and Ni-$d_{z^2}$ orbitals $V$ increase when the Pr-doping concentration increases. The trend of $t_\perp^x$, $t_\perp^z$, and $V$ affected by chemical pressure are in line with that by the external pressure [40]. Since the effective interlayer antiferromagnetic superexchange interaction $J_\perp$ can be estimate as $J_\perp = 4t_\perp^2/U$ [44, 58]. The increment of $t_\perp^z$ would enhance the strength of $J_\perp$, which would tunes the conventional single-layer $d$-wave superconducting state to the $s$-wave pair symmetry and enhance the interlayer superconductor order [58] in $La_3Ni_2O_7$. Besides, the $V$ would enhance the superfluid stiffness and boost phase coherence of Cooper pairs [44, 47], and its increment may be benefited for the forming of Cooper pairs. Therefore, Pr-doped $La_3Ni_2O_7$ would in favour of the superconducting state and reduce the pressure of the superconductivity obtained in $La_3Ni_2O_7$. While the intralayer nearest-neighbor hopping $t_1^x$ decreases with the Pr-doping concentration increasing. Such depression of $t_1^x$ reflect a more localized nature of in-plane Ni-$d_{x^2-y^2}$ orbitals. And this phenomenon reflected that the effect of chemical pressure on the itinerant nature of in-plane Ni-$d_{x^2-y^2}$ orbitals exhibit a contrary tendency compared with external pressure.

At 15 GPa, the $t_1^x$, $t_\perp^x$, $t_\perp^z$, and $V$ increase with the Pr-doping concentration increasing. The enhancement of hopping parameters partially reflects a relative weaker correlation of Pr-doped than that of pure $La_3Ni_2O_7$. And it means that the chemical pressure plays a similar role on the electronic properties of *I*4/*mmm* phase of $La_3Ni_2O_7$ compared with that of external pressure [40]. The value of $V/t_1^x$ is suggested to have a close relationship with $T_c$ in previous study [46]. This value slight increases from 0.47 of *I*4/*mmm*-$La_3Ni_2O_7$ to 0.48 of *P*4*mm*-$La_{11}PrNi_8O_{28}$. Therefore, doping Pr into *I*4/*mmm*-$La_3Ni_2O_7$ would enhance the superconductivity of $La_3Ni_2O_7$ and may explained the slight enhancement of onset $T_c$ in high pressure experimental [52].

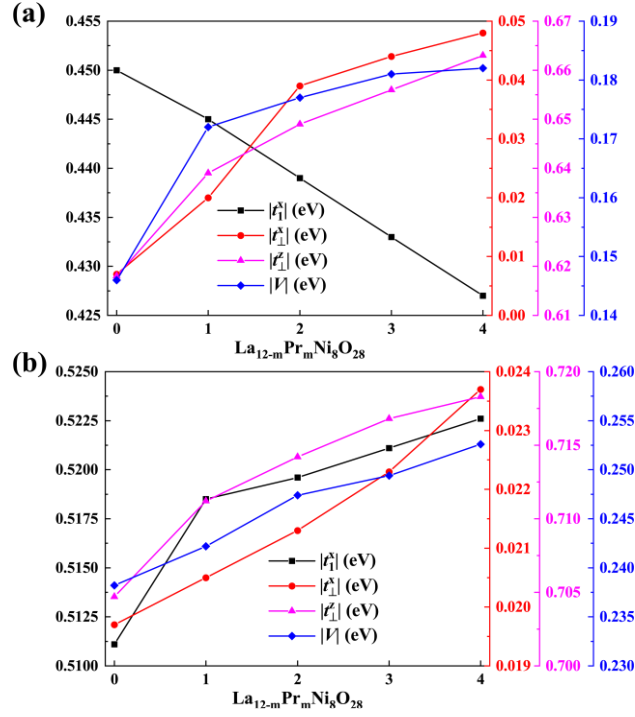

FIG. 5. Calculated part of the tight-binding parameters of the bilayer two-orbital model as a function of Pr-doping concentration at (a) ambient pressure and (b) 15 GPa.

In summary, we performed a comprehensive first-principles study on the crystal structures and electronic properties of Pr-doped $La_3Ni_2O_7$. Our results show that the praseodymium atoms preferentially occupy the outer La-O layer site. Besides, we suggested that evolutionary trends of chemical pressure for ambient pressure and high pressure phase with doping concentration are distinct. Analysis of electronic structures indicates that doping Pr into *Amam* phase of $La_3Ni_2O_7$ would enhance the interlayer coupling strength of Ni-$d_{x^2-y^2}$ orbitals and shifts the two proximal flat bands of Ni-$d_{z^2}$ orbitals near the $\Gamma$ point upward the $E_F$. Based on these findings, we explore the relationship between the Pr doping concentration and superconductivity, suggesting that Pr doping may promote superconductivity, as supported by our results of bilayer two-orbital model. This work provided valuable insights into the effect of chemical pressure in isovalent doped RP phase nickelate superconductor and offers a foundation for further exploration in this field.

We thank Jinguang Cheng for the experimental data and discussion. This work was supported by National Natural Science Foundation of China (Grants Nos. 12274169,

12122405, and 52072188), National Key R&D Program of China (Grants Nos. 2022YFA1402304 and 2023YFA1406200), Fundamental Research Funds for the Central Universities (Grants No. xzy022023011 and xhj032021014-04), and Program for Science and Technology Innovation Team in Zhejiang (Grant No. 2021R01004). Some of the calculations were performed at the High Performance Computing Center of Jilin University and using TianHe-1(A) at the National Supercomputer Center in Tianjin.